\documentclass[12pt]{iopart} 
 \usepackage{amssymb}
\usepackage{graphicx}
\usepackage{lipsum}
\usepackage{iopams}

\expandafter\let\csname equation*\endcsname\relax

\expandafter\let\csname endequation*\endcsname\relax

\usepackage{amsmath}
\usepackage{bbold}
\usepackage{bm}
\usepackage{caption}

\newcommand{\HH}{\mathcal{H}}

\begin{document}
       \title[Bond Order via Light-Induced Synthetic Many-body Interactions]{Bond Order via Light-Induced Synthetic Many-body Interactions of Ultracold Atoms in Optical Lattices}
        \author{Santiago F.  Caballero-Benitez$^{1,2*}$ and Igor B. Mekhov$^1$}
        \ead{$^*$scaballero@fisica.unam.mx}
      \address{ 
 $^1$University of Oxford, Department of Physics, Clarendon Laboratory, Parks Road, Oxford OX1 3PU, UK}
 \address{
 $^2$Instituto de F\'\i sica, Universidad Nacional Aut\'onoma de M\'exico, Apartado Postal 20-364, M\'exico DF 01000, Mexico}

\begin{abstract}
We show how bond order emerges due to light mediated synthetic interactions in  ultracold atoms in optical lattices in an optical cavity. This is a consequence of the competition between both short- and long-range interactions designed by choosing the optical geometry. Light induces effective many-body interactions that modify the landscape of quantum phases supported by the typical Bose-Hubbard model. Using exact diagonalization of small system sizes in one dimension, we present the many-body quantum phases the system can support via the interplay between the density and bond (or matter-wave coherence) interactions. We find numerical evidence to support that dimer phases due to bond order are analogous to valence bond states. Different possibilities of light-induced atomic interactions are considered that go beyond the typical atomic system with dipolar and other intrinsic interactions. This will broaden the Hamiltonian toolbox available for quantum simulation of condensed matter physics via atomic systems. 

\end{abstract}

       \maketitle

\section{Introduction}

Ultracold gases loaded in optical lattices are an ideal tool for studying competing phases of quantum matter. Engineering the effective potential seen by the atoms using light beams allows to realize with optical lattices simple models of condensed matter, particle physics and  even biological systems~\cite{Lewenstein}. Moreover,  the  realization of these models in experiments would aid in the development of applications towards quantum information processing (QIP) and the development of novel quantum materials via quantum simulation~\cite{Bloch}. Typically, one can realize effective Hamiltonians which contain short-range physical processes such as tunneling between neighbor lattice sites and on-site interactions according to a prescribed lattice potential engineered by classical light fields, such as the Bose-Hubbard (BH) model. Long range interactions are experimentally challenging but accessible in principle via  polar molecules~\cite{lahaye2009,Ferlaino} or Rydberg atoms~\cite{pohl2009,PohlLukin,Fleischhauer}. However, the nature of the interaction is fixed by the characteristics of its constituents. In addition, finite range interactions by other approaches of light-matter interactions~\cite{Porras2006,IonsFR2012,Lesanovsky2013,PhotCrys2015}, and extended Bose-Hubbard models via dipolar interactions~\cite{EBHZoller,EBHLewenstein} are  also possible.

In contrast to the above, loading an optical lattice inside a cavity allows to engineer effective synthetic many-body interactions between light induced atomic modes with an arbitrary spatial profile~\cite{PRL2015,NJPhys2015,PRA2016}. These interactions are mediated by the light field and do not depend on the nature of the atoms considered, making them extremely tunable and suitable for realizing quantum simulations of effective many-body long-range Hamiltonians. In principle fermionic, bosonic, molecular systems, etc. can be studied. This allows to explore the interplay between additional non-conventional quantum many-body phases, besides from the typical superfluid (SF) and Mott-insulator (MI). It is now experimentally possible to access the regime where light-matter coupling is strong enough with cavity decay rates of MHz~\cite{PNASEsslinger2013, Esslinger2015} and kHz~\cite{PNASHemmerich2015, Hemmerich2015}, in the range to compete with typical short-range processes (tunneling and on-site interactions).  Moreover, bosonic ultracold atoms loaded in an optical lattice inside an optical cavity have been recently realized~\cite{Hemmerich2015,Esslinger2015}, opening a new venue to analyse the interplay between competing orders of quantum matter by design. The light inside the cavity can be used to control the formation of many-body phases of matter even in a single cavity mode by properly choosing the arrangement of the cavity, optical lattice and light pumped into the system~\cite{EPJD08,MekhovRev,PRL2015, NJPhys2015}. Additional freedom can be achieved by multimode cavities or multiple cavities extending the possibilities to condense into exotic quantum phases even further~\cite{Strack2011,KramerRitschPRA2014,Muller2012,PRA2016}. Recent advances~\cite{LevNaturePhys, Kollar2015,Esslinger2}, will enable the experimental realization of synthetic interactions by design with additional freedom in the near future. This can be realized as the cavity parameters (decay rates) and detunnings with respect the cavity modes can be externally modified with respect to the atomic system. Moreover, the spatial profile of the cavity modes can be designed depending on the geometry of the coherent light beams pumped into the high-Q cavity~\cite{PRA2016}.

\begin{figure}[t!]
\centering
\includegraphics[width=0.75\textwidth]{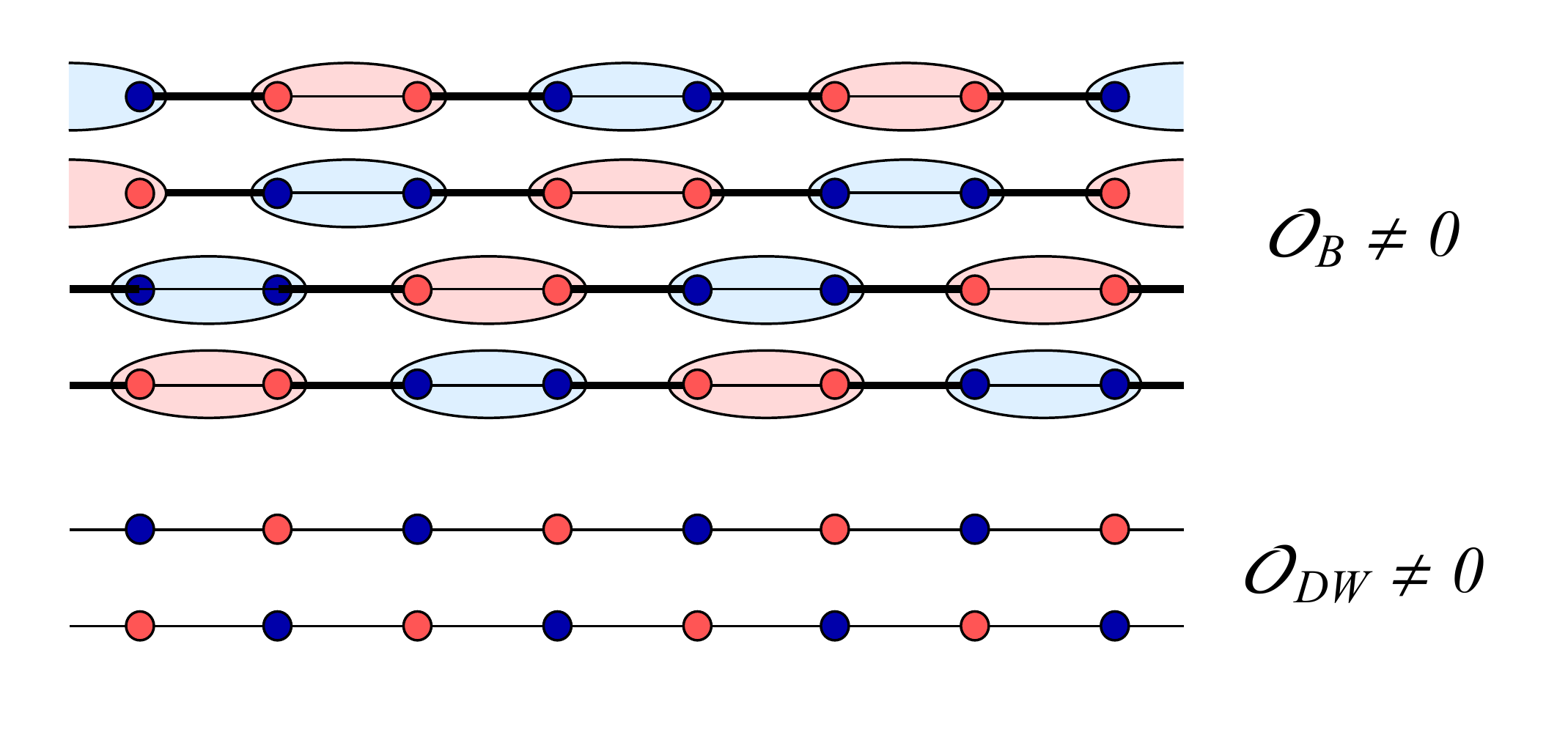}
\captionsetup{width=0.75\textwidth,justification=centerlast,font=small}
\caption{Typical configurations with bond order $\mathcal{O}_B$ and density wave order $\mathcal{O}_{DW}$ of the ground state, circles denote sites $i$. We discuss the competition between these orders in the main text and their definition. (Top) For typical $\mathcal{O}_B\neq0$, different ellipsoids correspond to pairs of nearest neighbor sites with the same matter field phase (dimers), colors denote different values of 
$\langle\hat b^\dagger_i\hat b^{\;}_{i+n}+\mathrm{h.c.}\rangle=c_{i}$ with a constant $c_i>0$, $n>i$. The thick lines in between ellipsoids have $\langle\hat b^\dagger_{i+1}\hat b^{\;}_{i+n+1}+\mathrm{h.c.}\rangle=\tilde c_i$, with $\tilde{c}_i\neq c_i$ adjacent dimers. If $c_{i+n+1}=-c_i$ with $\tilde{c_i}=0$ then maximal phase difference is stablished. if $\tilde c_i=0$ and $c_i\neq0$ maximal matterwave amplitude difference occurs. See main text on the structure of the one-body reduced density matrix.  
(Bottom) Different colors correspond to density values at sites $i$ for $\mathcal{O}_{DW}\neq0$. For homogenous states, superfluid and Mott insulator there is no pattern in coherences or densities, $\tilde{c}_i=c_i\forall n$.  The ground state is 4-fold degenerate for $\mathcal{O}_B$ and 2-fold degenerate for $\mathcal{O}_{DW}$.   
}\label{GS}
\end{figure}

In this article we present how different arrangements involving multiple probes and/or multiple light modes configurations, lead to the competition of atypical quantum many-body phases via synthetic interactions. In the past, single cavity competition between typical SF and MI phases~\cite{Larson,Morigi}, density wave orders~\cite{Hofstetter,Reza,PRL2015,NJPhys2015,PRA2016,Zhai2016,Donner2016}, and disorder~\cite{Morigi2} has been addressed. In contrast to other works, here we consider the interplay with what we call ``bond order'' and other orders in the system. Bond order is a form of self-organization~\cite{RitschRMP} of matter-wave coherences or ``bonds" due to cavity backaction to compensate for the phase difference imposed by the pattern of light pumped into the cavity and scattered by the  atoms~\cite{PRL2015}. In the effective Hamiltonian, as the number of cavity or pump modes increases different physics are possible due to the light induced atomic mode structure. Thus, competition of quantum many-body phases triggers due to the induced atomic mode structure (breaking symmetries, e.g. time reversal and translational) and the regular BH Hamiltonian processes ( homogenous tunnelling and on-site interactions). Particularly difficult is the regime where strong correlations will be present in addition to the well known physics of the BH Hamiltonian. This occurs when there is a  large effective light-matter coupling relative to on-site interactions. In this limit, standard mean-field theory becomes unreliable as strong imbalanced configurations can occur and large on-site fluctuations take place due to the broken symmetry of the ground state~\cite{PRA2016}. As the interaction has a global character with non-trivial structure, and in fact is of infinite range, symmetries are broken even in 1D. The ground state can acquire states with different competing orders, Fig.\ref{GS}. Thus, quantum phase transitions different from the usual type in 1D systems, of the  Berezinsky-Kosterlitz-Thouless (BKT) type, can occur.  In what follows, we analyse the system by using exact diagonalization for small number of sites in 1D. Our simulations are an indicative picture of the expected behaviour in a larger system. Additionally, we will study the behaviour related with the competition between between bond-order and other orders present in the system. In particular, we find numerical evidence to support the analogy between valence bond states (VBS)~\cite{AKTL,Tasaki}  and dimerised states that arise in the cavity system due to bond ordering via different mechanisms. Our results could be used as basis for the quantum simulation of analogous dimer states important in quantum magnetism~\cite{Balents}. Using classical optical lattices the AKLT Hamiltonian can be implemented in principle as the large interaction limit of the Bose-Hubbard Hamiltonian~\cite{Delgado1}. Non-trivial entanglement properties have been found~\cite{Delgado2, Delgado3}, while bulk-boundary duality with entangled pair states occurs~\cite{Pollmann, CiracBB} and spin glasses  are also possible~\cite{Sglass}. The states are potentially useful for measurement based quantum computation~\cite{Doherty}. Other ultracold systems where the possibility of bond ordered states has been explored include dipolar gases with nearest neighbour and truncated finite range density interactions\cite{dip1,dip2}, in the framework of the so-called extended Bose-Hubbard models~\cite{Ferlaino,EBHLewenstein}. Bond order can also occur via density dependent frustration of the hopping amplitudes with Raman assisted tunelling~\cite{SantosBO}. 
Moreover in condensed matter systems, bond order was first explored in low dimensions via extended Hubbard models with two species fermions~\cite{EHM1,EHM2,EHM3,EHM4}. 

 In our treatment, we provide an alternative route to explore physics of this kind using cavity fields that relaxes the constraint on very strong on-site interaction ($U$),  as cavity coupling can be tailored to simulate effective Hamiltonians with similar properties. In ultracold atoms in the fermionic version of our system, a closer analogy to investigate and simulate resonance valence bond (RVB) states important in high-Tc superconductivity~\cite{AndersonRVB}, might be possible.  Moreover, the competition between superconductivity and density wave orders is actively studied~\cite{Hawthorn1, Hawthorn2}, and light induced superconductivity is being researched~\cite{LISC1,LISC2,LISC3}.  We analyse the emergence and competition between superfluid, supersolid, insulating and dimerised quantum many-body phases of matter by means of the behaviour in their order parameters.

Our findings, will foster the study of competing orders in multicomponent optomechanical systems~\cite{RMP2014Optomech}. Moreover, the interplay of the quantum phases we study and their generalization, may appear in hybrid system networks~\cite{QnetCirac,QnetBeige,QnetRempe,QnetViola}.  In connection to our work, non interacting fermions in  cavity systems have been studied \cite{Fermions1, Fermions2, Fermions3} and even chiral states have been found~\cite{CFermions}.  Towards quantum state engineering via measurement back-action, competition with other correlated  quantum many-body states~\cite{GabrielAFM,PRA2009,LP09,LP10,LP11,LesanovskyDis, ZimmermanRing,Gabriel,GabrielPRA,Molmer,Vuletic,RitschArxiv2015,Kozlowski2016} and Non-Hermitian dynamics~\cite{Diehl,TELee,Dhar,WojciechNHQZ} can occur. Moreover, feedback control~\cite{Zimmerman,Pedersen,Denis1,Denis2,Denis3,Denis4,Feedback} can also be explored in relation to the dynamical stabilization of quantum many-body phases. 
As such, the behaviour of the emergent phases in the cavity system we will show,  might aid towards the design of novel quantum materials with analogous properties. It follows, that the use of the mechanisms described here could be incorporated in the future development of real materials and composite devices in hybrid solid state systems~\cite{Hybrid}, where both light and matter are in the quantum limit and quantum coherence can be exploited.

The article has the structure that follows. We introduce the general model of ultracold atoms in high-Q cavity(ies) where the atoms are in the regime of quantum degeneracy. We continue by stating the effective models we will consider due to synthetic interactions between light induced atomic spatial modes. Then, we define the different competing orders that can arise in the system. Next, we present our results via the phase diagrams of competing phases. Finally, we conclude our manuscript by summarising our findings.

\section{The model}
The system consists of atoms trapped in an OL inside a cavity (single/multi mode) or many cavities with the cavity mode frequency(ies) $\omega_c$ and decay rate(s) $\kappa_c$~\cite{MekhovRev,Gabriel,Wojciech,PRL2015,NJPhys2015,PRA2016,Atoms2015,Thomas}. The atomic system is subject to additional light beam(s) pumped into the system in off-resonant light scattering. {The off-resonant light scattering condition means that $\Gamma \ll |\Delta_{pa}|$, where $\Gamma$ is  the spontaneous emission rate of the atoms, where  $\Delta_{pa}=\omega_p-\omega_a$ is the detuning between the light mode(s)  frequency(ies) $\omega_p$ and the atomic resonance frequency $\omega_a$.} The scattered light from the ultracold atoms in the OL is selected and coupling is enhanced by the optical cavity(ies), generating a quantum potential. The light pumped into the system has amplitude(s)  $\Omega_p\in\mathbb{C}$ (in units of the Rabi frequency). The pump-cavity detunnings are $\Delta_{pc}=\omega_p-\omega_c$. The light is pumped from the side of the main axis of the high-Q cavity(ies), at an angle not necessarily at $90^\circ$ which allows for arbitrary control of the overlap of  light induced spatial modes~\cite{PRA2016}.  The cavity modes couple with the atoms via the effective coupling strengths $g_p= g_c \Omega_p/(2\Delta_{pa})$, with $g_c$ the light-matter coupling coefficient of the cavity. The light-matter Hamiltonian describing the system after the light-field has been adiabatically eliminated~\cite{EPJD08,NJPhys2015} in the good cavity limit  ($\kappa_c\ll\Delta_{pc}$) is:
 $\HH_{\mathrm{eff}}=\HH^b+\HH^{ad}$, where $\HH^b$ is the regular Bose-Hubbard (BH) Hamiltonian~\cite{Fisher,Dieter},
\begin{equation}
\HH^b=-t_0\sum_{\langle i, j\rangle}(\hat b^\dagger_i\hat b^{\phantom{\dagger}}_j+h.c)-\mu\sum_i\hat n_i+\frac{U}{2}\sum_i\hat n_i(\hat n_i-1),
\end{equation}
with $t_0$ the nearest neighbour tunneling amplitude, $U$ the on-site interaction and $\mu$ the chemical potential. The operators $b_i^\dagger$ ($\hat b_i$) create (annihilate) bosonic atoms at site $i$,  the number operator of atoms per site is given by $\hat n_i=\hat b_i^\dagger\hat b_i^{\phantom{\dagger}}$.  The on-site interaction and hopping amplitude terms are short-range local processes. The BH Hamiltonian contains the effective parameters forming the classical optical lattice~\cite{Dieter}.
The emergent effective light-induced interaction is~\cite{NJPhys2015,PRA2016}, 
\begin{equation}
\HH^{ad}=\sum_{\varphi,\varphi'}\sum_c\sum_{p,q}\big(\tilde\gamma^{D,D}_{\varphi,\varphi'}(c,p,q)\hat N_\varphi^{\phantom{*}}
\hat N_{\varphi'}^{\phantom{*}}+\tilde\gamma^{B,B}_{\varphi,\varphi'}(c,p,q)\hat S_\varphi^{\phantom{*}}\hat S_{\varphi'}^{\phantom{*}}
+\tilde\gamma^{D,B}_{\varphi,\varphi'}(c,p,q)[\hat N_\varphi^{\phantom{*}}\hat S_{\varphi'}^{\phantom{*}}
+\hat S_{\varphi'}^{\phantom{*}}\hat N_{\varphi}^{\phantom{*}}]
\big),
\label{Had}
\end{equation}
where 
\begin{eqnarray} 
\tilde\gamma^{\eta,\nu}_{\varphi,\varphi'}(c,p,q)&=&\frac{|\tilde g_{pc}|^2}{2}\left(\frac{(J_{\eta,\varphi}^{pc})^* J^{qc}_{\nu,{\varphi'}}}{\Delta_{qc}+i\kappa_c}+c.c.\right),
\label{cstr}
\end{eqnarray}
with $\tilde g_{pc}=g_c\Omega_p/(2\Delta_{pa})$ where $\{\eta,\nu\}\in\{D,B\}$. The sum over ``$p$" ans ``$q$" go over the number of pumps and ``$c$" goes over the cavity modes (for a multi-mode cavity/several cavities). The couplings $J^{pc}_{\eta,\varphi}\in\mathbb{C}$, correspond to the possible values of $J^{pc}_{ij}$ (Wannier overlap integrals)~\cite{PRA2016, NJPhys2015,PRL2015, Gabriel, Thomas,Atoms2015,Wojciech} for each mode of the cavity system through the inter-site amplitudes, labeled $B$, or  through the site density, labeled $D$. These can either be for a single mode cavity with one pump and one cavity or  a multi-mode cavity, and even multiple cavities and multiple pumps. These coupling constants are given by,
  \begin{equation} 
J^{pc}_{ij}=\int w(\mathbf{x}-\mathbf{x}_i)u^*_p(\mathbf{x})u_c(\mathbf{x})w(\mathbf{x}-\mathbf{x}_j)\mathrm{d}^n x,
\end{equation}
where ``$i$" and "$j$" can be the same site  for density coupling or be nearest neighbours for bond coupling (inter-site densities), where $u_{c,p}(\mathbf{x})$ are the cavity(ies) and pump(s) mode functions, typically travelling or standing waves. The $w(\mathbf{x})$ are the Wannier functions given by the classical optical lattice in the lowest band. The light induced ``density"  $\hat N_\varphi$ and  ``bond"  $\hat S_{\varphi}$ mode operators are such that:
\begin{equation}
\hat N_\varphi=\sum_{i\in\varphi}\hat n_i ,\; \textrm{and}\; \hat S_{\varphi}=\sum_{\langle i,j\rangle\in\varphi}(\hat b_i^\dagger\hat b_j^{\phantom{\dagger}}+\hat b_j^\dagger\hat b_i^{\phantom{\dagger}}),
\end{equation}
The sums go over illuminated sites $N_s$ and nearest neighbour pairs $\langle i,j\rangle$ that belong to the light-induced atomic spatial mode $\varphi$. As it has been shown~\cite{PRA2016} the coupling constants can be designed with great freedom by choosing the angle of incident light with respect to the classical optical lattice plane and the cavity axis. The spatial structure of light is useful as a natural basis to define these atomic modes, as the coupling coefficients $J^{pc}_{ij}$ can periodically repeat in space~\cite{PRL2015,NJPhys2015,PRA2016, Gabriel, GabrielPRA,Thomas,Wojciech}. The atoms that belong to a particular light-induced atomic mode scatter light with the same phase. Thus, one can use the distribution of values of $J^{pc}_{ij}$, to define the light induced spatial atomic modes. As the pump and cavity modes are external to the internal structure of the system (the BH model), they provide a large set of independently tuneable parameters. This allows to tailor the effective light-induced atomic mode interaction with an arbitrary spatial profile. By addressing the density via the couplings $J^{pc}_{D,\varphi}$ one can generate multi-component density orders.  Density wave orders correspond to different groups of atoms for each light induced atomic mode. In the case of $J^{pc}_{B,\varphi}$ one can generate dimer, trimers, tetramers, etc. which will form as a consequence of the pattern induced to the matter-wave coherences or ``bonds". These two kinds of orders will compete in addition with the superfluid order in the system and the Mott insulating phase of the BH model as we will see. The current effective model disregards additional density dependent Wannier functions modified dynamically by light, which are difficult to calculate self-consistently. However the proper redefinition and self-consistent determination of these functions won't alter the essential structure of the effective Hamiltonian. This will only renormalise the effective coupling strengths and parameters of the Bose Hubbard model. Thus our results are applicable in a frame of reference with this renormalized parameters.  In addition, coupling between cavity modes has not been included, as these processes have much smaller amplitudes compared to the pump modes.

\section{Effective Hamiltonians}

In contrast to previous works, here we focus in the large effective light matter interaction where quantum fluctuations cannot be accounted for in mean-field theory regarding bond order. Moreover, we will consider the interplay between density coupling and bond order  in the strong-coupling limit. To do this we will analyse the following Hamiltonian corresponding to a single cavity and a single pump, where the incident light illuminating from the side has been designed to scatter through the bonds and densities as a staggered field (at $90^\circ$ with respect to the cavity axis~\cite{PRL2015,PRA2016,Wojciech}) with components effectively tuned by the couplings of the bond $J_B$ and densities $J_D$, 
\begin{equation}
\HH_{\mathrm{eff}}=\HH^b+\frac{g_{\mathrm{eff}}}{N_s}\left[J_B^2\hat B_-^2
+J_D^2\hat D_-^2+J_BJ_D(\hat B_-\hat D_-+\hat D_-\hat B_-)\right],
\label{JDJBH}
\end{equation}
the effective interaction strength is $g_{\mathrm{eff}}/N_s\sim\tilde{\gamma}=\Delta_c|\tilde g|^2/(\Delta_c^2+\kappa_c^2)$, which depends on amplitude of the light pumped into the system and the light detunnings, Eq.(\ref{cstr}). The  bond $\hat B_-$ and density operators $\hat D_-$ are:
\begin{eqnarray}
\hat B_{-}=
\sum_{i=0}^{N_s-1}(-1)^{i}(\hat b_i^\dagger\hat b_{i+1}^{\phantom{\dagger}}+\mathrm{h.c.}) 
\quad\textrm{and}\quad
\hat D_{-}=
\sum_{i=0}^{N_s-1}(-1)^{i}\hat n_i
\end{eqnarray}
For ultracold atoms in an optical cavity in the adiabatic limit, cavity decay rates are of the order of MHz. The effective interaction strength, $g_{\mathrm{eff}}$, can be typically be made of the same order of magnitude  or larger than on-site interactions, $|g_{\mathrm{eff}}|\gtrsim U\sim t_0\sim E_R\sim\textrm{kHz}$~\cite{Esslinger2015}, with $E_R$ the recoil energy. Note that the ratio $t_0/U$ can be tuned via the classical optical lattice depth and/or Feshbach resonances\cite{Dieter}. Essentially the sign of the light induced interaction can be chosen via the cavity-pump detunning $\Delta_c$ and the amplitude by the pump strength $\Omega_p$\cite{PRL2015,NJPhys2015,PRA2016}.   In addition, without loss of generality, $\{J_B,J_D\}\in[0,1]$. Depending on the lattice depth of the classical optical lattice, e.g. the Bose-Hubbard Wannier functions and the choice of illumination, the magnitude of the $J_{B,D}$ coupling constants can be tuned using real Wannier functions~\cite{Wojciech}. Beyond a gaussian ansatz this gives $J_B\neq0$ depending on the lattice depth of the classical optical lattice. Typically for a lattice depth of $5 E_R$ (where the single band approximation is valid) then $J_B\approx 0.05|\sin[(\Sigma\phi-\Delta\phi)/2]\cos[(\Sigma\phi+\Delta\phi)/2]|$ while $J_D\approx 0.8|\cos[(\Sigma\phi-\Delta\phi)/2]\cos[(\Sigma\phi+\Delta\phi)/2]|$ with $\Delta\phi=\phi_1-\phi_0$ and $\Sigma\phi=\phi_1+\phi_0$.   $\Delta\phi$ ($\Sigma\phi$) is the difference (sum) of phases between two crossed standing waves (with phases $\phi_{0,1}$) pumped from the side at $90^\circ$ with respect to the classical optical lattice potential. The beams are arranged such that $k_{0x}=0$ and $k_{x,1}=\pi/a$, with $a$ the lattice spacing (typically $a=\lambda/2$ for a standing wave in the classical optical lattice). Therefore, the ratio between the value of the two contributions can be adjusted arbitrarily, e.g for $J_B/J_D=0.25$ we have $\Delta\phi\approx0.844\pi$ with $\Sigma\phi=0$ for simplicity. Thus, any ratio between the coefficients $J_D$ and $J_B$ is possible and can be modified in addition by changing the classical optical lattice depth, maximal bond coupling is achieved by $\Delta\phi=\pi/2$ while maximal density coupling is achieved by $\Delta\phi=0$ with $\Sigma\phi=0$. By increasing the depth of the classical optical lattice the coefficient $J_B$ becomes smaller and it is basically negligible for a lattice depth of 15 $E_R$. 

\begin{figure}[t!]
\centering
\includegraphics[width=0.85\textwidth]{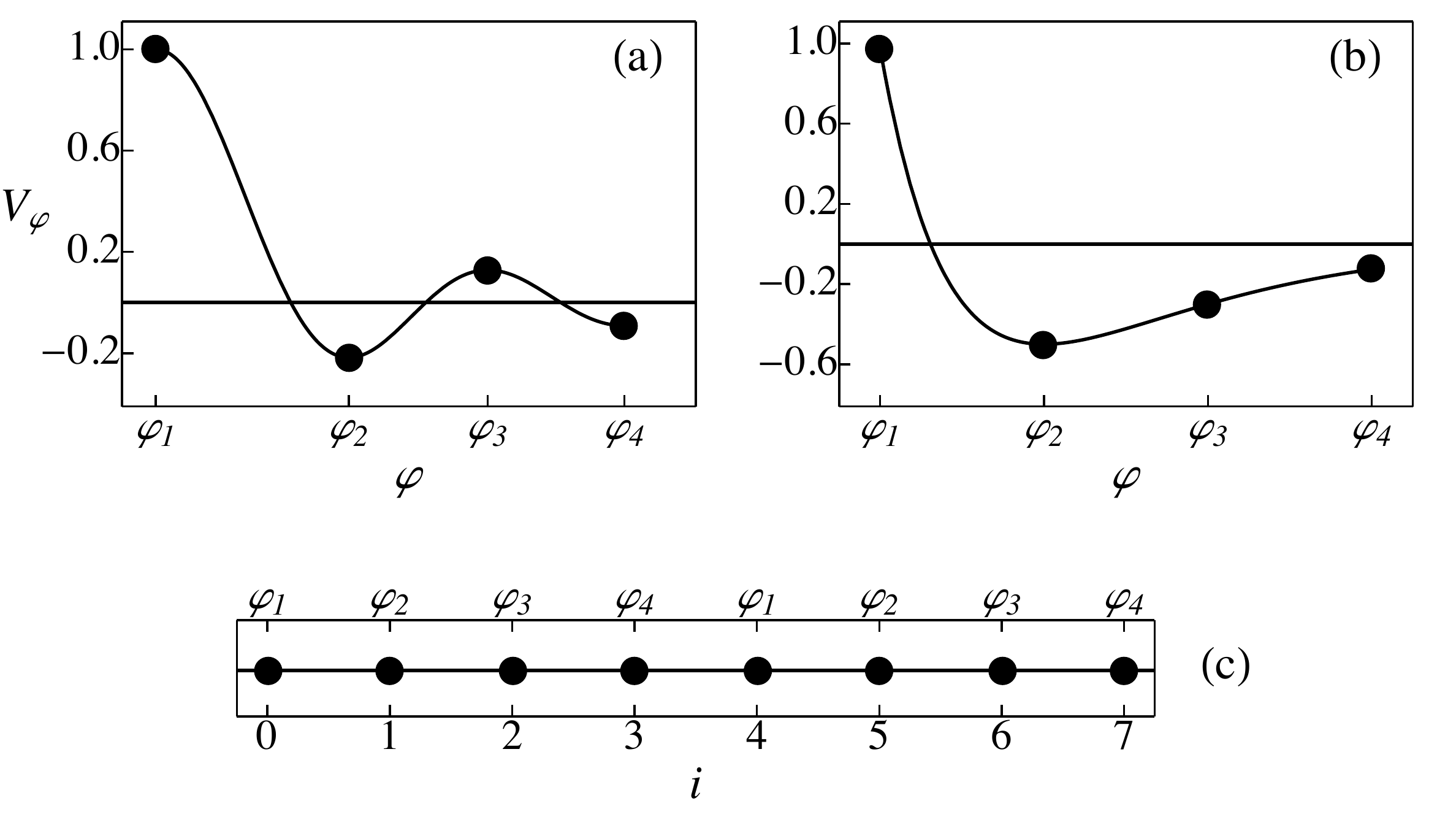}
\captionsetup{width=0.85\textwidth,justification=centerlast,font=small}
\caption{Effective light induced mode interactions  $V_\varphi$. (a) Bessel type potential, the mode coupling constants have been chosen for the first $R=4$ minima and maxima. (b) Morse type potential. (c) The correspondence rule between lattice points $i$ and  the light-induced modes $\varphi_m$,  with $m=1,2,3,4$ for the mode coupling coefficients in (a) and (b).}\label{PotFig}
\end{figure}
Moreover, we consider the previously unexplored scenario where via multiple cavities or multiple pumps one can perform the quantum simulation of the following light induced atomic mode interactions~\cite{PRA2016}:
 \begin{itemize}
 \item{Bessel type potential, $V_{\varphi_m}= j_0(\pi(x_m-1))$, see Fig.\ref{PotFig}(a).}
\item{Morse type potential, $V_{\varphi_m}=\big[\big(1-e^{-(m-2)}\big)^2-1\big]/2$, see Fig.\ref{PotFig}(b).}
\end{itemize}
The function $j_0$ is the zero order spherical Bessel function of the first kind and $\varphi_m$ with $m\in[1,R]$ with $R$ light-induced spatial atomic modes. $x_m-1$ are the locations of the maxima and minima of $j_0(y)$ with $y\in[0,3]$.   The maximum amplitude of the interactions has been chosen such that, $\max(V_{\varphi_m})\sim1$, for simplicity.  Morse type potentials are  a typical phenomenological tool to model effective molecular systems. The Bessel type potential we consider, shows an example of the flexibility of the construction with respect to the degree of control that can be achieved via the synthetic light-induced atomic mode interactions. The relationship with the pump and cavity coupling via inverse discrete Fourier transforms can be found in general in \cite{PRA2016}. Certainly other types of potentials can be tailored with great flexibility depending on the quantum many-body system we would like to simulate. 
 For many cavity modes (multiple cavities/ multimode cavity) we have,
\begin{equation}
\HH_{\mathrm{eff}}=\HH^b+\frac{g_{\mathrm{eff}}}{ N_s}\sum_{\varphi,\varphi'}V_{|\varphi-\varphi'|}\hat N_\varphi\hat N_{\varphi'},
\label{s1H}
\end{equation}
where the interaction depends on the mode distance $|\varphi-\varphi'|$. This kind of effective many-body interaction is physically motivated to account for finite range effective interacting potentials. For many pumps in a single mode cavity, we have,
\begin{equation}
\HH_{\mathrm{eff}}=\HH^b+\frac{g_{\mathrm{eff}}}{N_s}\sum_{\varphi,\varphi'}V_{\varphi}V_{\varphi'}\hat N_\varphi\hat N_{\varphi'}
\label{s2H}
\end{equation}
where the interaction depends on the position between light-induced atomic modes. This is potentially useful for the simulation of biological systems and other hybrid networks~\cite{QnetCirac,QnetBeige,QnetRempe,QnetViola}. Here in contrast to the many cavity mode case, the interaction is position dependent and corresponds to the interaction between different branches, channels or nodes in the network.  Without loss of generality, we will consider the case of $R=4$ light-induced atomic modes, such that $\varphi_m$ with $m\in[1,4]$ for simplicity. The correspondence rules between light induced modes $\varphi_m$ and lattice sites $i$ is shown in Fig.\ref{PotFig}(c) for 1D  lattice with $N_s=8$. In principle, the number of pump modes can be arbitrarily increased by shining the light at different angles with respect to the cavity axis in combination with beam splitters. We call our interactions synthetic, as they are artificially designed by the choice of the spatial profile of the light pumped into the system and the cavity modes~\cite{PRA2016}. The properties of the light pumped into the system and the cavity parameters are external to the intrinsic properties of the atoms ($t_0$ and $U$) and easily tuned in the range of the atomic processes  of the order of the recoil energy.

\section{Order Parameters}

Bond order occurs whenever dimerized structures appear in the ground state of the Hamiltonian. These bosonic dimerised structures, akin to valence bond states (VBS)~\cite{AKTL}, appear in the particular case where the structure of light-matter coupling alternates sign in the inter-site amplitudes or bond between two neighbouring sites~\cite{PRL2015}. Concretely, the ground state of the system is such that in order to maximise light scattering the inter-site coherences self-organise to minimise the energy. This can be extracted from the ground state configuration via a function of the operator $\hat B_-$.  In exact diagonalisation, if  $\langle\hat B_-\rangle=0$, it inherently implies a degenerate quantum superposition (Schr\"odinger ``cat-state"), if $\langle\hat B_-^2\rangle\neq0$. Thus, a useful order parameter regarding bond-order can be defined as,
\begin{equation}
\mathcal{O}_B^2
=\frac{\langle\hat B_-^2\rangle}{N_s^2},
\end{equation}
akin to a staggered magnetisation, a bond order structure factor~\cite{SantosBO}. In the case when $\mathcal{O}_B\neq0$ there is imbalance between mater wave coherences in the ground state. This is a manifestation of a broken time reversal symmetry in the ground state.  This is not necessarily coexisting with broken translational invariance, e. g. a ground state with density wave order. It is worth noting that this order parameter will signal bond-order whenever we are not in a MI~\cite{Wojciech}. Deep in the  MI  with exact diagonalization, we have: $\mathcal{O}_B^2|_{\textrm{MI}}=2(\rho+1)\rho/N_s$ with $\rho\in\mathbb{Z}^+$, which in large $N_s$ limit vanishes. For a density wave insulator with maximal imbalance we have: $\mathcal{O}_B^2|_{\textrm{DW}}=2\rho/N_s$. In order to tell the difference between a  MI and a bond ordered state we will use the fact that the on-site fluctuations $\Delta(n_i)^2=\langle\hat n_i^2\rangle-\rho^2$ are zero in the MI. Thus,  we have bond order when $\Delta(n_i)\neq$ and $\mathcal{O}_B\neq 0$. When bond order emerges matter-wave coherence patterns can be from a slight imbalance between matter-wave coherences, e.g. $\langle\hat b^\dagger_n\hat b^{\phantom{\dagger}}_{n+1}+\mathrm{H.c.}\rangle\neq \langle\hat b^\dagger_{n+1}\hat b^{\phantom{\dagger}}_{n+2}+\mathrm{H.c.}\rangle$. Maximal phase difference  in the coherences is stablished when:  $\langle\hat b^\dagger_n\hat b^{\phantom{\dagger}}_{n+1}+\mathrm{H.c.}\rangle=-\langle\hat b^\dagger_{n+1}\hat b^{\phantom{\dagger}}_{n+2}+\mathrm{H.c.}\rangle$.
Maximal matterwave coherence amplitude difference is stablished when $\langle\hat b^\dagger_n\hat b^{\phantom{\dagger}}_{n+1}+\mathrm{H.c.}\rangle\neq0$ and $\langle\hat b^\dagger_{n+1}\hat b^{\phantom{\dagger}}_{n+2}+\mathrm{H.c.}\rangle=0$. The typical matter-wave (MW) coherence patterns found with bond order in the one-body reduced density matrix of sites $(i,j)$ are:

\begin{itemize}
\item{Partial MW amplitude imbalance:}
$$
\left(
\begin{array}{cccccccc}
 \rho _0 & \tilde{c} & c & \tilde{c} & c & \tilde{c} & c & \tilde{c} \\
 \tilde{c} & \rho _1 & \tilde{c} & c & \tilde{c} & c & \tilde{c} & c \\
 c & \tilde{c} & \rho _2 & \tilde{c} & c & \tilde{c} & c & \tilde{c} \\
 \tilde{c} & c & \tilde{c} & \rho _3 & \tilde{c} & c & \tilde{c} & c \\
 c & \tilde{c} & c & \tilde{c} & \rho _4 & \tilde{c} & c & \tilde{c} \\
 \tilde{c} & c & \tilde{c} & c & \tilde{c} & \rho _5 & \tilde{c} & c \\
 c & \tilde{c} & c & \tilde{c} & c & \tilde{c} & \rho _6 & \tilde{c} \\
 \tilde{c} & c & \tilde{c} & c & \tilde{c} & c & \tilde{c} & \rho _7 \\
\end{array}
\right)
$$
\\
\item{Maximal MW amplitude imbalance:}
$$
\left(
\begin{array}{cccccccc}
 \rho _0 & 0 & c & 0 & c & 0 & c & 0 \\
 0 & \rho _1 & 0 & c & 0 & c & 0 & c \\
 c & 0 & \rho _2 & 0 & c & 0 & c & 0 \\
 0 & c & 0 & \rho _3 & 0 & c & 0 & c \\
 c & 0 & c & 0 & \rho _4 & 0 & c & 0 \\
 0 & c & 0 & c & 0 & \rho _5 & 0 & c \\
 c & 0 & c & 0 & c & 0 & \rho _6 & 0 \\
 0 & c & 0 & c & 0 & c & 0 & \rho _7 \\
\end{array}
\right)$$\\
\item{Maximal MW  phase difference imbalance:}
$$
\left(
\begin{array}{cccccccc}
 \rho _0 & 0 & -\tilde{c} & 0 & c & 0 & -\tilde{c} & 0 \\
 0 & \rho _1 & 0 & -\tilde{c} & 0 & c & 0 & -\tilde{c} \\
 -\tilde{c} & 0 & \rho _2 & 0 & -\tilde{c} & 0 & c & 0 \\
 0 & -\tilde{c} & 0 & \rho _3 & 0 & -\tilde{c} & 0 & c \\
 c & 0 & -\tilde{c} & 0 & \rho _4 & 0 & -\tilde{c} & 0 \\
 0 & c & 0 & -\tilde{c} & 0 & \rho _5 & 0 & -\tilde{c} \\
 -\tilde{c} & 0 & c & 0 & -\tilde{c} & 0 & \rho _6 & 0 \\
 0 & -\tilde{c} & 0 & c & 0 & -\tilde{c} & 0 & \rho _7 \\
\end{array}
\right)$$
\end{itemize}
for  $\tilde{c}$ and $c$ positive real constants, where each entry in the matrix  corresponds to $\langle\hat b^\dagger_i\hat b^{\phantom{\dagger}}_{j}\rangle$ for $\{i,j\}\in 0,\dots,N_s-1$. Thus, distant matter-wave amplitudes are correlated. In a perfect SF ($U=0$, e.g. $g_{\mathrm{eff}}=0$), $c=\tilde{c}=\rho_i=\rho$. Deep in the MI ($t_0=0$, e.g. $g_{\mathrm{eff}}=0$) $\rho_i=\rho\in\mathbb{Z}^+$, $c=\tilde{c}=0$. A pictorial representation is given in Fig.\ref{GS}. On the other hand the matrix representing, the product of nearest neighbour coherences can be constructed, a typical structure for bond-ordered states with maximal phase difference between MW is the following:
$$
\left(
\begin{array}{cccccccc}
 \alpha  & -\beta  & \tilde{\lambda } & -\lambda  & \tilde{\lambda } & -\lambda  &
   \tilde{\lambda } & -\lambda  \\
 -\beta  & \alpha  & -\beta  & \tilde{\lambda } & -\lambda  & \tilde{\lambda } &
   -\lambda  & \tilde{\lambda } \\
 \tilde{\lambda } & -\beta  & \alpha  & -\beta  & \tilde{\lambda } & -\lambda  &
   \tilde{\lambda } & -\lambda  \\
 -\lambda  & \tilde{\lambda } & -\beta  & \alpha  & -\beta  & \tilde{\lambda } &
   -\lambda  & \tilde{\lambda } \\
 \tilde{\lambda } & -\lambda  & \tilde{\lambda } & -\beta  & \alpha  & -\beta  &
   \tilde{\lambda } & -\lambda  \\
 -\lambda  & \tilde{\lambda } & -\lambda  & \tilde{\lambda } & -\beta  & \alpha  &
   -\beta  & \tilde{\lambda } \\
 \tilde{\lambda } & -\lambda  & \tilde{\lambda } & -\lambda  & \tilde{\lambda } & -\beta
    & \alpha  & -\beta  \\
 -\lambda  & \tilde{\lambda } & -\lambda  & \tilde{\lambda } & -\lambda  &
   \tilde{\lambda } & -\beta  & \alpha  \\
\end{array}
\right)
$$
 where $\alpha$,$\beta$,$\lambda$, $\tilde\lambda$ are positive real constants.  Here each entry corresponds to the product of elements $\langle\hat s_n\hat s_m\rangle$ with $\hat s_m=(\hat b^\dagger_m\hat b^{\phantom{\dagger}}_{m+1}+\mathrm{H.c.})$. The alternating character of the sign of its elements is characteristic of bond ordered states, e.g. for SF and SS all elements are positive. Note that $\mathcal{O}_{B}^2=(1/N_s^2)\sum_{n,m}(-1)^{n+m}\langle\hat s_n\hat s_m\rangle$ (where we have used periodic boundary conditions, e.g. $\hat b_{N_s}=\hat b_0$). In the large $N_s$ limit we have for a bond ordered state: $\mathcal{O}_{B}^2|_{\mathrm{BO}}\approx\lambda$ where $\lambda\approx-\tilde\lambda>0$. In terms of the above matrix elements, deep in the MI or DW insulators we have $\lambda=\tilde\lambda=\beta\approx0$, thus  $\mathcal{O}_{B}^2|_{\mathrm{MI/DW}}=\alpha/N_s$ for $N_s\gg1$. 

Moreover, the above order will compete and coexist with density wave order, typically given by the structure factor~\cite{Batrouni},
\begin{eqnarray}
\mathcal{O}_{DW}^2=\frac{1}{N_s^2}\sum_{i,j}(-1)^{|i-j|}\langle\hat n_i\hat n_j\rangle
\equiv\frac{\langle\hat D_-^2\rangle}{N_s^2}.
\end{eqnarray}
DW order breaks translation invariance in the ground state and signals a $\mathbb{Z}_2$ symmetry between odd and even sites (Schr\"odinger ``cat state" in the density configurations). Deep in a density wave insulator the bond order parameter is: $\mathcal{O}_B^2|_{\textrm{DW}}=2\rho/N_s$ while $\mathcal{O}_{DW}^2=\rho^2$ is maximal.

We use the condensate fraction as an estimator of the SF fraction in the system,
\begin{equation}
f_{SF}\sim f_c=\frac{1}{2N_s}\sum_{i=0}^{N_s-1}\langle\hat b_i^\dagger\hat b_{i+1}^{\phantom{\dagger}}+\mathrm{h.c.}\rangle.
\end{equation}
 Alternatively one could use the difference in energy with respect to a phase twist~\cite{Burnett}. For an ideal SF, $f_{SF}=\rho$.

At commensurate fillings, in addition to a MI the system can present hidden string order, the string order parameter is given by,
\begin{equation}
\mathcal{O}_S=\lim_{|i-j|\to\infty}\langle\delta\hat n_ie^{i\theta\sum_{i\leq k<j}\delta\hat n_k}\delta\hat n_j\rangle
\end{equation}
with $\delta\hat n_k=\hat n_k-\rho$ and $\rho$ the average density per-site (the filling factor). In order to distinguish the MI and string ordered states, it is necessary to define the parity order parameter,
\begin{equation}
\mathcal{O}_P=\lim_{|i-j|\to\infty}\langle e^{i\theta\sum_{i\leq k<j}\delta\hat n_k}\rangle
\end{equation}
In combination with the other order parameters, the string and parity order parameters allow to distinguish the emergence of a Haldane insulator (HI)~\cite{Tasaki,Altman,Batrouni}. While $\mathcal{O}_{DW}=0$, $\mathcal{O}_P=0$, and $f_{SF}=0$,  if $\mathcal{O}_S>0$,  with a gapped spectrum, the system is a HI. If the spectrum is gapless ($f_{SF}\neq0$) or gapped  (insulator, $f_{SF}=0$) and $\mathcal{O}_S>0$,  $\mathcal{O}_P=0$, then we have a type of VBS, a dimerised phase. For filling $\rho=1$, $\theta=\pi$, otherwise $\theta$ needs to be determined with the help of additional methods~\cite{theta}.

If the system is in the SF state with coexisting bond order  $\mathcal{O}_{B}\neq0$, it is in the superfluid dimer phase (SFD). The system has matter wave coherence patterns but is homogenous in the density. The typical difference between the order in the ground state for either $\mathcal{O}_B=\sqrt{|\mathcal{O}_B^2|}$ or $\mathcal{O}_{DW}=\sqrt{|\mathcal{O}_{DW}^2|}$ is shown in Fig.\ref{GS}. Note that whenever $\mathcal{O}_{DW}>\mathcal{O}_{B}$ the system will be in a DW phase either an insulator if the SF component is zero or a supersolid phase (SS) if $f_{SF}\neq0$ . If $\mathcal{O}_{DW}<\mathcal{O}_{B}$ and with SF component different from zero the system will be in a supersolid dimer phase (SSD). In the SSD, the system has density variation and matter wave coherence pattern with finite superfluid fraction. Whenever the system is in the SS,  SSD or SFD phases, the spectrum is gapless as there is a finite SF component in the system. In addition, it can occur that the system is in a bond insulator (BI) phase, where $\mathcal{O}_{DW}=0$ and $\mathcal{O}_{B}>0$ and the SF component is zero at incommensurate fillings. Here the system, has phase pattern but there is no SF fraction or DW order and it is not a MI. In BI, dimerised structures form the ground state and it is homogenous in space.  It can also happen that, we have a coexistence insulating phase where, $0<\mathcal{O}_{DW}\leq\mathcal{O}_B$, while the superfluid fraction is completely suppressed, this is a BI+DW insulating phase. This phase is a dimer insulator akin to VBS with density imbalance between components.  For commensurate filling $\rho=1$ the BI+DW phase presents $\mathcal{O}_S\neq 0$ and $\mathcal{O}_P=0$. Thus bond ordered phases can be gapped (BI) and gapless (SFD, SSD). SSD and SFD phases are bear similarities of RVB states\cite{AndersonRVB}, being  bosonic gapless  ground states with dimerised structures. On the other hand, BI states are similar to VBS, being gapped.

As finite size effects are considerable for small number of sites in the order parameters, to circumvent this problem, we have used the fact that in the large $t_0/U$ limit the system will tend to be a perfect SF. Thus, all other order parameters besides $f_{SF}$ should approach zero. Therefore, the finite size spurious contribution in other order parameters is eliminated by renormalising with respect to the perfect SF value ($f_{SF}$ in the limit $t_0/U\gg0$).  We subtract the SF fraction profile multiplied by the large $t_0/U$ limit off-set due to finite size. Besides from this, some intermediate phases found will be harder to observe as the number of sites increases, concretely: SFD, SSD and SS which appear as the system moves from insulating states when $t_0/U=0$ to the ideal SF in the limit $t_0/U\gg1$.

In table \ref{T1}, we summarise the quantum many-body phases of the system and the relation with the order parameters defined.

 In what follows, we will constrain our discussion on the half-filled  $\rho=1/2$, and integer filling $\rho=1$ cases, while considering simulations for $N_s=8$ and renormalized order parameters as previously explained.

\begin{table}
\begin{center}
\begin{tabular}{r|rrrrrr}
QP       & $\mathcal{O}_B$  &$\mathcal{O}_{DW}$ &$\mathcal{O}_{P}$& $\mathcal{O}_{S}$& $\Delta(\hat n_i)$ & $f_{SF}$\\
\hline
SF               &  0       & 0          &0          & 0          &$\neq0$ &$\neq0$\\
SS               &   0     &$\neq0$&0           &$\neq0$ &$\neq0$ &$\neq0$\\
SSD            &$\neq0$     &$\neq0$&0           &$\neq0$ &$\neq0$ &$\neq0$\\
SFD            &$\neq0$ &0            &0           &$\neq0$ &$\neq0$ &$\neq0$\\
MI               & 0        & 0          &$\neq0$&0            &0            &0\\ 
DW             & 0&$\neq0$&0           &$\neq0$ &$\neq0$&0 \\
BI               &$\neq0$&0         &0            &$\neq0$ &$\neq0$&0 \\
BI+DW       &$\neq0$&$\neq0$&0          &$\neq0$ &$\neq0$&0\\
\end{tabular}
\end{center}
\caption{
Relation between order parameters and quantum many-body phases (QP). The criteria to distinguish in our finite size simulations has been relaxed to define DW, SS with $\mathcal{O}_{DW}>\mathcal{O}_B$ and SFD with $\mathcal{O}_{B}\gg\mathcal{O}_{DW}$. The discussion on several finite size effects is in the main text.}
\label{T1}
\end{table}

\section{Results}

\subsection{Bond order vs Density wave order }
\begin{figure*}[t!]
\centering
\includegraphics[width=0.85\textwidth]{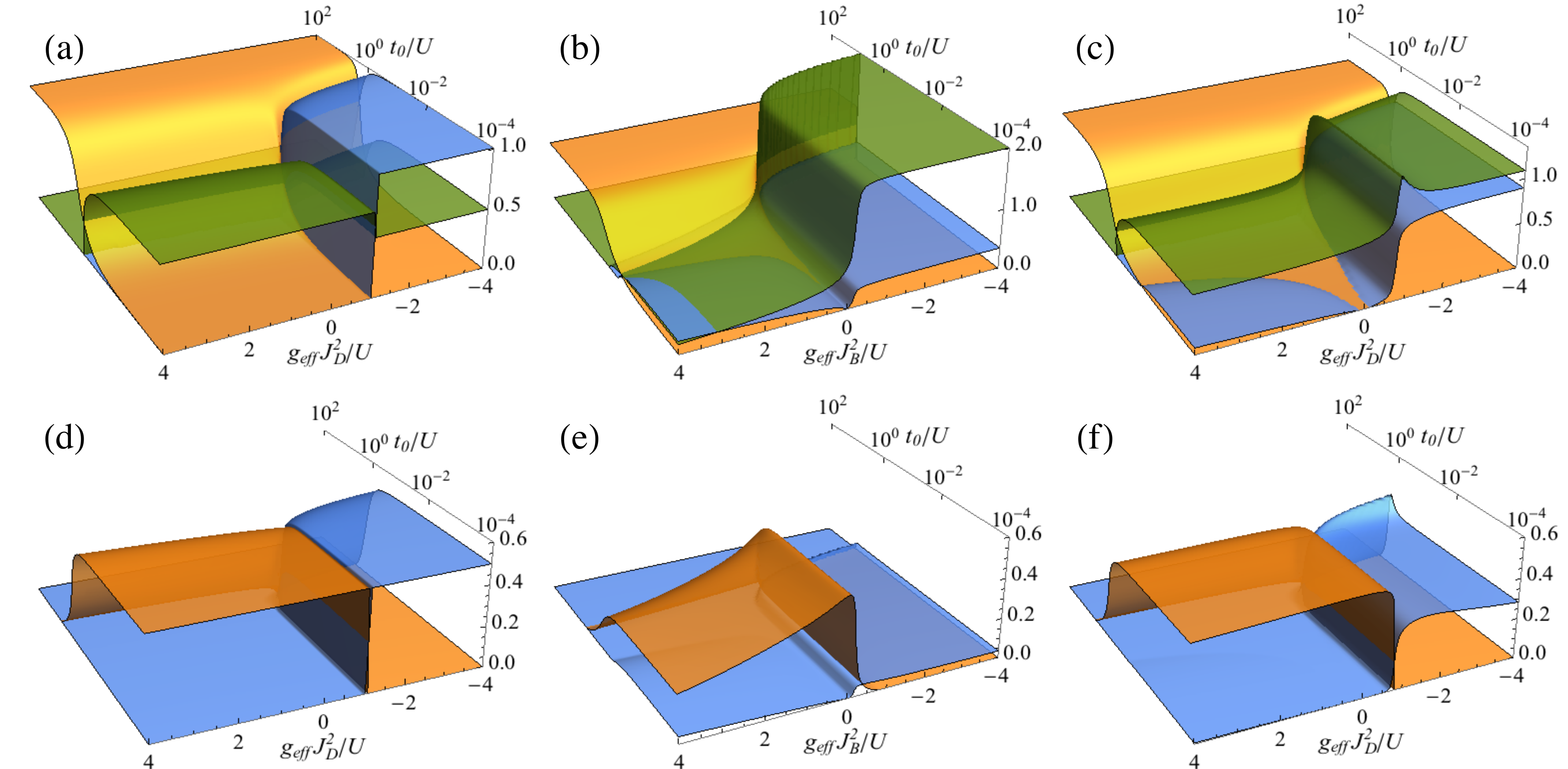}
\includegraphics[width=0.85\textwidth]{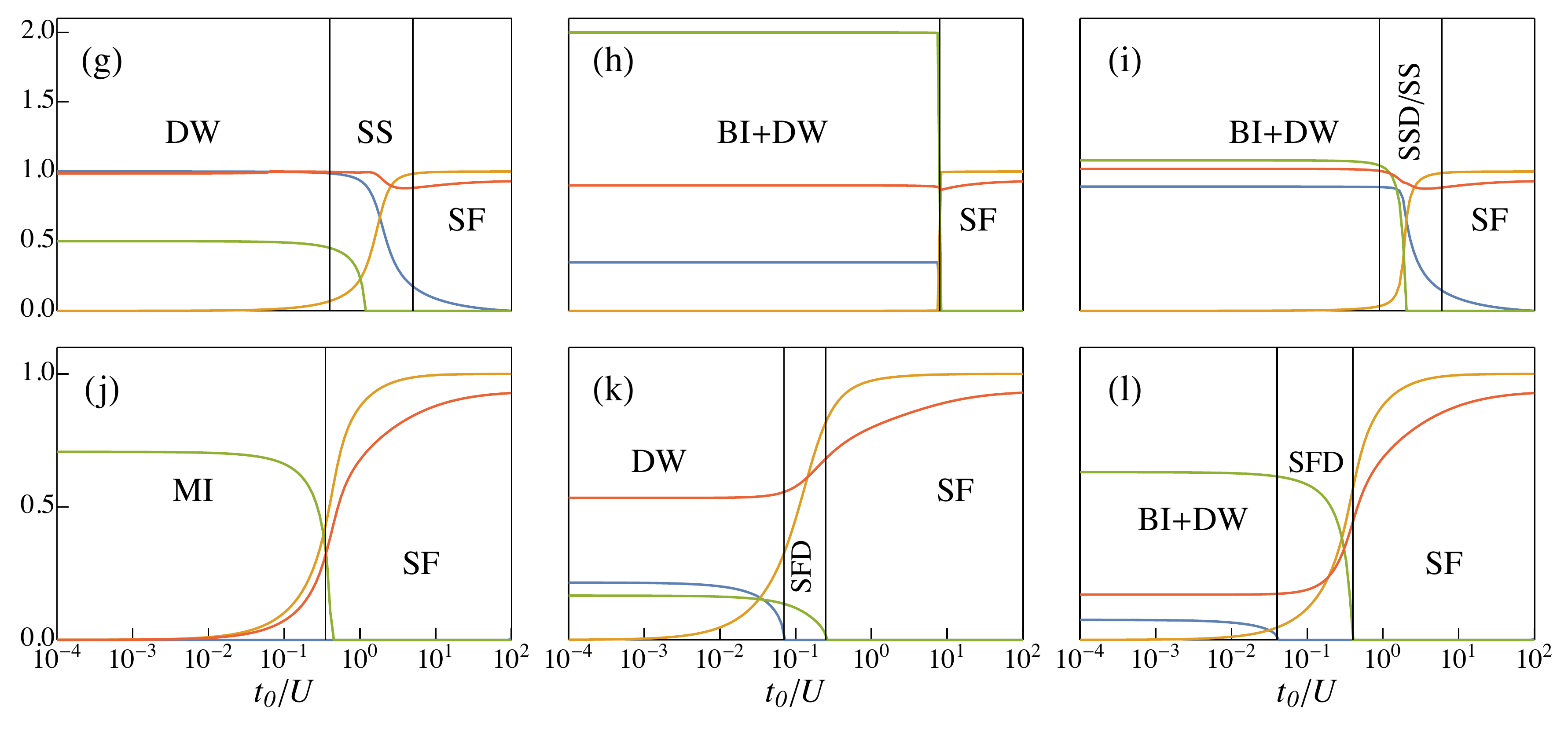}
\captionsetup{width=0.85\textwidth,justification=centerlast,font=small}
\caption{Order parameters at integer filling $\rho=1$ for $J_{B,\varphi}=\pm J_B$ and $J_{D,\varphi}=\pm J_D$ as a function of the tunneling $t_0/U$ and $g_{\mathrm{eff}}$ in units of $U$. Panels (a-c): $f_{SF}$ SF fraction (yellow), $\mathcal{O}_B$ bond order parameter (green), $\mathcal{O}_{DW}$ density wave order parameter (blue). Panels (d-e): $\mathcal{O}_P$ parity order parameter (blue) and $|\mathcal{O}_S|$ string order parameter (orange). Parameters in (a), (d), (g), and (j) show the system with $J_D\neq0$ and $J_B=0$. The SF-MI transition gets shifted with respect to its $g_{\mathrm{eff}}J_D^2=0$ value. Below $g_{\mathrm{eff}}J_D^2<-U$, DW, SS and SF phases are supported by the system. Parameters in (b), (e), (h), and (k) show $J_B\neq0$ and $J_D=0$. For large on-site interactions ($t_0/U$ small) the system supports DW for $g_{\mathrm{eff}}J_B^2>0$ and BI+DW for $g_{\mathrm{eff}}J_B^2<0$. Parameters in (c), (f), (i), (l) show $J_D\neq 0$ and $J_B\neq0$ with $J_B/J_D=0.25$, DW order and Bond order compete. DW insulator is supported for $g_{\mathrm{eff}}J_D^2>0$. Panels (g) to  (l):   $\mathcal{O}_B$ (green), $\mathcal{O}_{DW}$ (blue), $f_{SF}$ (yellow) and on-site fluctuations $\Delta(\hat n_i)$ (red). Panels (g),(h), and (i) correspond to $g_{\mathrm{eff}}J_{D,B}^2/U=-4$.  Panels (j),(k), and (l) correspond to $g_{\mathrm{eff}}J_{D,B}^2/U=+4$.
Parameters in all panels are: $N_s=8$ with $\rho=1$.
}
\label{JBJD1}
\end{figure*}

In this section we will analyse the results from simulations performed using the effective Hamiltonian (\ref{JDJBH}).

At integer filling $\rho=1$, the simplest case to understand is when there is only density coupling ($J_D\neq 0$ and $J_B=0$), as the density wave instability forms for negative $g_{\mathrm{eff}}$. The system in addition to SF and MI states is able to support the emergence of DW insulator and SS phases, see Fig.\ref{JBJD1} (a) and (d).
The SF and MI exist for $g_{\mathrm{eff}}J_D^2>-U$ while the transition point shifts to higher values of $t_0/U$ as $g_{\mathrm{eff}}J_D^2>0$ and smaller values for $g_{\mathrm{eff}}J_D^2<0$, compared to the system without cavity light.  This occurs as on-site fluctuations are enhanced because light scatters minimally being a quantum optical lattice effect, not recoverable by simple mean-field analysis as corrections must be included~\cite{PRL2015} .Once $g_{\mathrm{eff}}J_D^2<-U$, the system has a discontinuous transition to the DW insulator state. While increasing the  
effective tunneling, the system goes from DW through SS to SF smoothly. This can be seen in the behaviour of $f_{SF}$ and  $\mathcal{O}_{DW}$, Fig.\ref{JBJD1}(g). The insulating character of the MI is confirmed by the absence of string order parameter $\mathcal{O}_S=0$ away from the DW order phases, while having $\mathcal{O}_P\neq0$, Fig.\ref{JBJD1}(d), while the onsite fluctuations are also minimal Fig.\ref{JBJD1}(j).

\begin{figure*}[t!]
\centering
\includegraphics[width=0.95\textwidth]{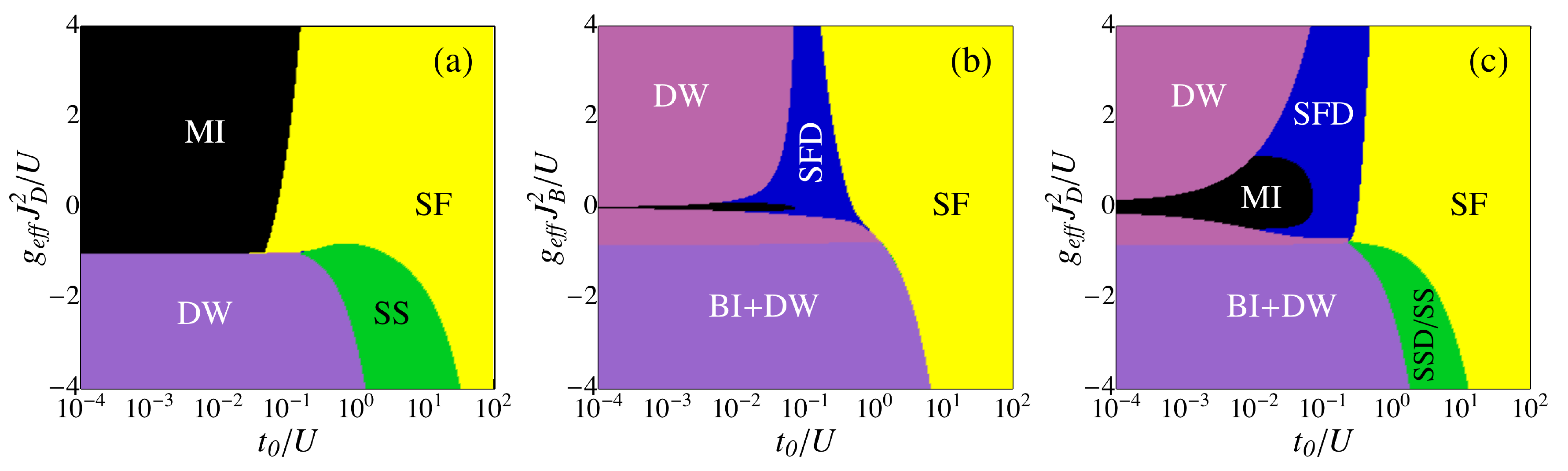}
\captionsetup{width=0.95\textwidth,justification=centerlast,font=small}
\caption{ Panel (a) shows the system with $J_D\neq0$ and $J_B=0$, bond order is not supported only DW (insulator and SS), MI and SF. Panel (b) shows $J_B\neq0$ and $J_D=0$, DW and BI insulators are supported. SFD phase exists as intermediate phase between DW and SF for $g_{\mathrm{eff}}J_B^2>0$. Panel (c) shows $J_B\neq0$ and $J_D\neq0$
with $J_B/J_D=0.25$, the intermediate SS in (a) turns into SSD and SS. The SSD has partial matter wave coherence imbalance between dimers. In (b) and (c) BI has maximal phase difference between dimers. The intermediate SS (a), SFD (b), and SSD/SS (c) phases shrink as the number of lattice sites $N_s$ increases, SF  takes over. Note that for $g_{\mathrm{eff}}=0$, the system only has MI and SF, the thin black region is not visible in (b) and (c).
Parameters in all panels are the same as in Fig. \ref{JBJD1}.
}
\label{PDrho1}.
\end{figure*}

When only dimer coupling occurs ($J_D=0$, $J_B\neq 0$), the system for strong on-site interactions $t_0/U\ll1$ supports DW for $g_{\mathrm{eff}}J_B^2\gg0$ and for  $g_{\mathrm{eff}}J_B^2<0$ BI+DW states. For  $g_{\mathrm{eff}}J_B^2>0$, the system evolves from DW to SF via an intermediate SFD phase as tunneling increases. As $g_{\mathrm{eff}}J_B^2\ll 0$ the system goes from the BI+DW state the SF phase as $t_0/U$ increases rather sharply, Fig.\ref{JBJD1}(h). Complementarily, Fig.\ref{JBJD1}(e), $\mathcal{O}_S$ is different from zero as SFD and BI+DW phases emerge. We have that $\mathcal{O}_B>\mathcal{O}_{DW}\neq0$ and $f_{SF}=0$, thus a BI+DW, a coexisting bond Insulating with density wave insulating phase. This state is different from a SSD state as there is no SF component (the state is gapped), and it is neither a MI, neither a HI. Summarising, the transitions from the insulating phases towards the SF state are sharp for  $g_{\mathrm{eff}}J_B^2>0$ and continuous for  $g_{\mathrm{eff}}J_B^2<0$. The  $\mathcal{O}_P$ smoothly decreases to zero as DW is approached for $g_{\mathrm{eff}}>0$, while there is sharp change as $g_{\mathrm{eff}}<0$ in the BI+DW phase, where $\mathcal{O}_P=0$. We consider this as numerical evidence to support that the BI+DW phase is analogous to a gapped VBS state.

The analogy between dimer states in the cavity system and VBS states can be traced back to the relationship between spin operators and bosonic operators via the Schwinger mapping~\cite{Auerbach}. Thus, the bond operators with alternating sign coupling in the bosonic system induce an analogous staggered field interaction. However, typically in spin systems interactions are of local character. In contrast to this, our interactions are global but not trivial, as they are structured.  This produces a similar mechanism for the formation of an anti-ferromagnetic like state. However, as we are considering soft-core bosons the analogy is not complete to spins. The dimers in the system are similar to the typical spin singlets of the original VBS~\cite{Tasaki}. The typical hardcore bosonic representation of the Bose Hubbard Hamiltonian (limit $U\to\infty$) via Matsubara-Matsuda mapping~\cite{MatMad} ($\hat b_i^\dagger\to S_i^+$, $\hat b_i^{\phantom{\dagger}}\to S_i^-$ and $\hat n_i\to S^z_i+1/2$) is:
\begin{equation}
\HH^b|_{\textrm{HC}}\approx-\frac{t_0}{2}\sum_i(S_i^+S_{i+1}^-+S_i^-S_{i+1}^+)-\mu\sum_i\hat S^z_i 
\end{equation}
which is an anisotropic Heisenberg model~\cite{Auerbach} while
\begin{equation}
\HH^{ad}|_{\textrm{HC}}\approx \frac{g_{\mathrm{eff}}}{N_s}\left(\sum_i\left[\frac{J_B}{2}(-1)^i(S_i^+S_{i+1}^-+S_i^-S_{i+1}^+)+J_D(-1)^i\hat S^z_i\right]\right)^2 
\end{equation}

\begin{figure*}[t!]
\centering
\includegraphics[width=0.9\textwidth]{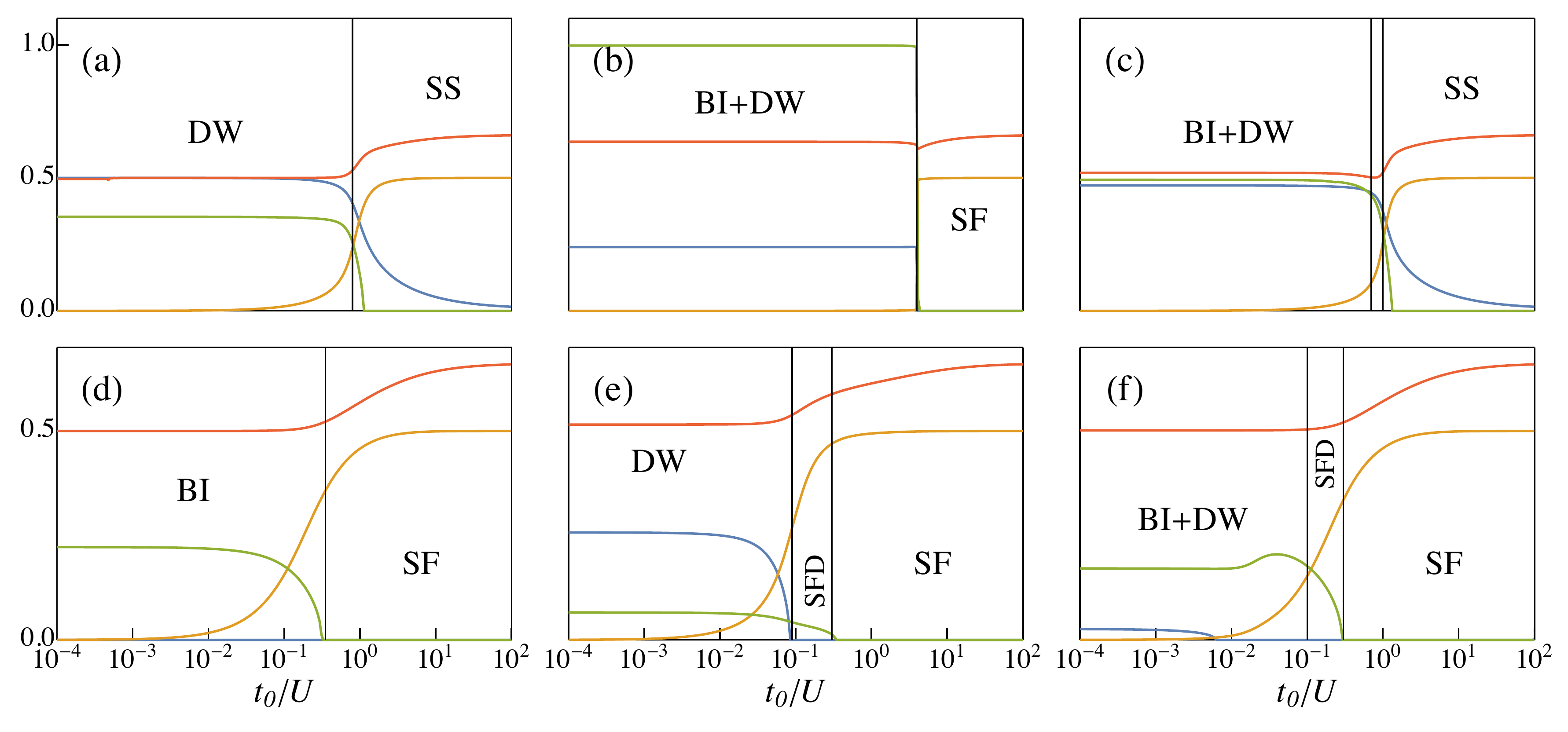}
\captionsetup{width=0.9\textwidth,justification=centerlast,font=small}
\caption{Order parameters at half filling $\rho=1/2$ for $J_{B,\varphi}=\pm J_B$ and $J_{D,\varphi}=\pm J_D$ as a function of the tunneling $t_0/U$. $f_{SF}$ superfluid fraction (yellow), $\mathcal{O}_B$ bond order parameter (green), $\mathcal{O}_{DW}$ density wave order parameter (blue), on-site fluctuations $\Delta(\hat n_i)$ (red). For $t_0/U\ll 1$, broken symmetry insulating states states emerge. Competition between DW and Bond ordered phases gives rise to stabilisation or suppression of DW order depending on the pump-cavity detunning via $g_{\mathrm{eff}}$. 
Parameters are: $N_s=8$ with $\rho=1/2$, (a) and (c): $g_{\mathrm{eff}}J_D^2/U=-4$, (b) $g_{\mathrm{eff}}J_B^2/U=-4$; (d) and (f): $g_{\mathrm{eff}}J_D^2/U=+4$, (e) $g_{\mathrm{eff}}J_B^2/U=+4$; (a) and (d) $J_B=0$, $J_D\neq0$; (b) and (e)  $J_B\neq0$, $J_D=0$; (c) and (f) $J_B\neq0$, $J_D\neq0$ with $J_B/J_D=0.25$. 
}
\label{JBJD05MM}
\end{figure*}

Now making the above contributions isotropic, only keeping the $J_B$ terms ($J_D=0$) and at fixed particle number  we have:
\begin{equation}
\HH^{\mathrm{eff}}|_{\textrm{HCS}}\approx-\frac{t_0}{2}\sum_i(\mathbf{S}_i\cdot \mathbf{S}_{i+1})+ \frac{g_{\mathrm{eff}} J_B^2}{4N_s}\left[\sum_i(-1)^i(\mathbf{S}_i\cdot \mathbf{S}_{i+1})\right]^2
\label{HCL}
\end{equation}
where $\mathbf{S}=(S^x,S^y,S^z)$. On the other hand the AKTL Hamiltonian~\cite{AKTL} is:
\begin{equation}
\HH^{\mathrm{AKLT}}=\alpha\sum_i(\mathbf{S}_i\cdot \mathbf{S}_{i+1})-\alpha\beta\sum_i(\mathbf{S}_i\cdot \mathbf{S}_{i+1})^2
\label{AKLT}
\end{equation}
Therefore, one can see (\ref{HCL}) is a global relative to (\ref{AKLT}) from which our Hamiltonian is an anisotropic relative without the hard core constraint formally. As such, some similarities might be expected between their ground states.

When density coupling and bond coupling act simultaneously ($J_B\neq0$ and $J_D\neq0$, with $J_B/J_D=0.25$), the situation interpolates in between the above two limits. However the additional bond-density coupling terms have strong effect even for small $J_B$. Bond ordering can take over the system behaviour instead of DW order, see Fig.\ref{JBJD1}(c), (f), (i) and (l). Interestingly, for large on-site interactions (small $t_0/U$) and $g_{\mathrm{eff}}J_D^2<0$, we find that a state with both DW and bond order occurs while being insulating, the DW does not destroy bond order.  When we increase the effective tunneling for  $g_{\mathrm{eff}}J_D^2\ll0$, the system smoothly transitions from the BI+DW state to the SF via a mixture of SSD and SS phases. In contrast, when $g_{\mathrm{eff}}>0$ and $t_0/U=0$ the system is a DW insulating state. The system transitions smoothly form this state one increases $t_0/U$ to the SF state via an intermediate SFD phase. In general, bond order takes over and competes with DW order as the ratio $J_B/J_D$ increases for strong on-site interactions while smoothly transitioning to the SF state as $t_0/U$ increases via  SSD and SS ($g_{\mathrm{eff}}J_D^2<0$) and superfluid dimer phases SFD ($g_{\mathrm{eff}}J_D^2>0$).
In the current parameter range explored there is no indication of a HI phase for $\rho=1$. The phase diagram of the above cases is shown in Fig.\ref{PDrho1}.

\begin{figure*}[t!]
\centering
\includegraphics[width=0.95\textwidth]{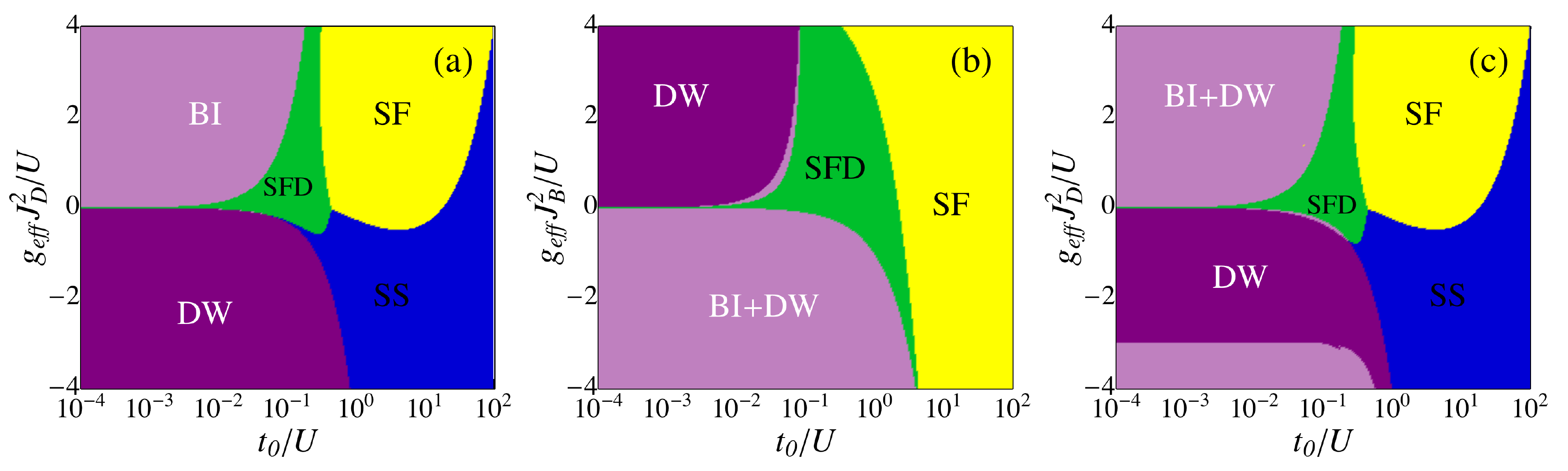}
\captionsetup{width=0.95\textwidth,justification=centerlast,font=small}
\caption{ Panel (a) shows the system with $J_D\neq0$ and $J_B=0$, BI, SFD, DW, SS and SF are supported. Bond ordered states (BI) and SFD occur with partial matterwave coherence imbalance. Panel (b) shows $J_B\neq0$ and $J_D=0$  BI with coexisting DW is supported. The BI+DW has maximal matterwave phase difference in dimers. Panel (c) shows $J_B\neq0$ and $J_D\neq0$
with $J_B/J_D=0.25$, the BI+DW insulator has maximal matter-wave coherence imbalance but no phase difference. The SFD in all panels has partial matter wave coherence imbalance between dimers.  The intermediate SFD phase shrinks as the number of lattice sites $N_s$ increases. SS states in (a) and (c) have small density imbalance.
Parameters in all panels are the same as in Fig. \ref{JBJD05MM}.
}
\label{PDrho05}.
\end{figure*}

It follows to consider the emergent phases at the half filled case ($\rho=1/2$). In the system without cavity light, we only have SF phase as there is no gap in the excitation spectra due to incommensuration for the homogenous system. However, even when $J_B=0$ and $J_D\neq0$, the induced symmetry breaking by light will foster the formation of insulators with broken translational  and time reversal symmetry. As function of the effective light matter strength for $t_0/U\ll1$ the system has a sharp transition from a DW insulator ($g_{\mathrm{eff}}J_D^2<0$) to a BI ($g_{\mathrm{eff}}J_D^2>0$). In the limit of $g_{\mathrm{eff}}J_D^2\ll0$, the system smoothly goes from a DW insulator to the SS phase, Fig.\ref{JBJD05MM}(a). In the opposite limit ($g_{\mathrm{eff}}\gg0$), the system for strong on-site interactions is a BI. As $t_0/U$ increases, the system smoothly reaches the SF state via an intermediate SFD phase, Fig.\ref{JBJD05MM}(d). Surprisingly, even with only density coupling, bond ordered phases arise in the large $U$ limit. This can be traced back to the fact that the one-body reduced density matrix has the structure where maximal amplitude MW coherence occurs in this case. This is a consequence of minimising the DW order and the fact that there is incommensuration.

Bond coupling ($J_B\neq 0$ and $J_D=0$) at $\rho=1/2$ has the effect of stabilising DW ordered phases (DW and SS) when $g_{\mathrm{eff}}J_B^2>0$. In contrast, the system supports  BI+DW phases for $g_{\mathrm{eff}}J_B^2<0$ and strong on-site interactions. Even at incommensurate fillings, and addressing through the bonds, the effect of cavity light is a suppression effect upon the SF component. This leads to have a sharp transition from the BI+DW phase to the SF state for   $g_{\mathrm{eff}}J_B^2\ll0$, Fig.\ref{JBJD05MM}(b). For  $g_{\mathrm{eff}}J_B^2\gg0$, the transition from the DW insulator to the SF state is smoothly connected via SS and later a SFD phase, Fig.\ref{JBJD05MM}(e).

Similar to the case with only bond coupling, bond ordered phases take over in the case of simultaneous addressing ($J_B/J_D=0.25$).  However, density coupling stabilises DW order, while BI+DW phases disappear for $g_{\mathrm{eff}}J_D^2\ll0$.The DW insulator takes over. The large $t_0/U$ limit phase for $ g_{\mathrm{eff}}<0$  is a SS and not a homogenous SF,  Fig.\ref{JBJD05MM}(c).The opposite effect occurs in the limit of $g_{\mathrm{eff}}J_D^2\gg0$, where instead DW ordered phases are strongly suppressed. The state of the system changes from BI+DW to SF via an intermediate SFD phase, Fig.\ref{JBJD05MM}(f). The phase diagram of the cases considered at half-filling is shown in Fig.\ref{PDrho05}.

It is expected that SS phases will shrink as the number of sites increases for Fig.\ref{PDrho05} (a) and (c). In general effects in the phase diagram when $g_\mathrm{eff}J_{D,B}^2>0$ are a manifestation of the quantum optical lattice potential induced by light to the atoms, as light is scattered minimally~\cite{PRL2015}. The insulating phases  (DW, MI) and intermediate phases that appear $g_\mathrm{eff}J_{D,B}^2>0$ are driven by quantum fluctuations.

\subsection{Bond order and Synthetic Potentials}
\begin{figure}[t!]
\centering
\includegraphics[width=0.75\textwidth]{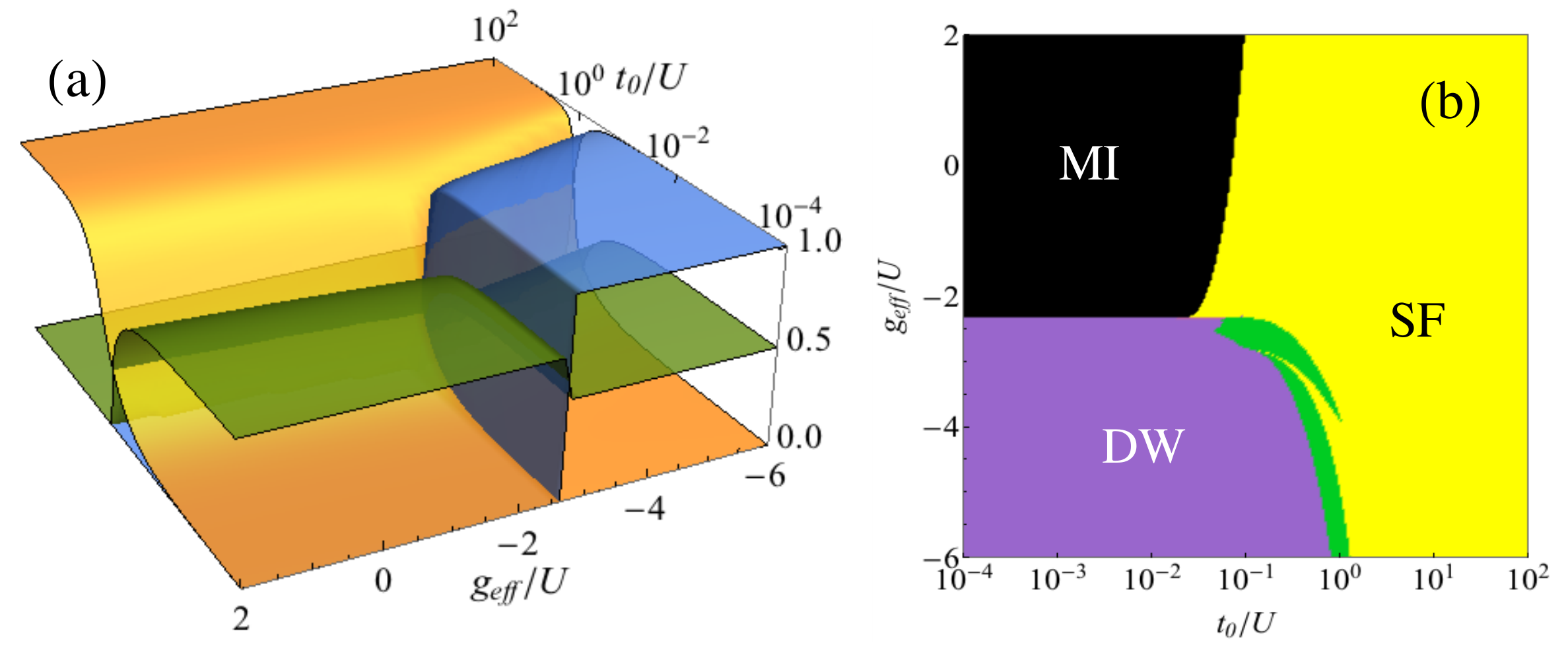}
\captionsetup{width=0.75\textwidth,justification=centerlast,font=small}
\caption{(a) Order parameters at integer filling $\rho=1$ for synthetic Bessel interaction with many cavity modes, as a function of the tunneling $t_0/U$ and $g_{\mathrm{eff}}/U$.  $f_{SF}$ superfluid fraction (yellow), $\mathcal{O}_B$ bond order parameter (green), $\mathcal{O}_{DW}$ density wave order parameter (blue). (b)  Phase diagram, the green region corresponds to SS. These are qualitatively similar for multiple cavity modes, multiple pumps and the single cavity single pump case but with a shifted critical $g_{\mathrm{eff}}$ for the emergence of DW order.  
Parameters are: $N_s=8$ with $\rho=1$ and $R=4$ light-induced modes.
}
\label{Bcav1}
\end{figure}

In this section we will study the results from simulations of effective Hamiltonians (\ref{s1H}) and (\ref{s2H}).  In the case where synthetic interactions via density coupling are considered, the situation is qualitatively the same for filling $\rho=1$ and either for many pumps  (\ref{s1H}) or many cavities  (\ref{s2H}) and either Bessel or Morse like potentials. There are slight changes in critical points  but overall the phase diagram can be summarised qualitatively in the phase diagram for Bessel-like interactions with many cavities, Fig.\ref{Bcav1}. For light-induced many-body effective interaction strengths such that $g_{\mathrm{eff}}>g_c$, with $g_c\approx-2.25U$ the system is in the MI for small $t_0/U$. The critical effective tunneling $t_0/U$ of the SF-MI transition shifts to lower values with respect to the system without cavity light for $g_{\mathrm{eff}}<0$, while for   $g_{\mathrm{eff}}>0$ the MI is stabilised to larger critical $t_0/U$. The situation is very similar to the density coupling  case in diffraction minima ($J_B=0$, $J_D\neq 0$), with a relevant shift in $g_c$ from $g_c\approx-U$ to $g_c\approx-2.25 U$. This can be attributed to the additional mode dependency of the synthetic interaction with $R=4$ modes. The additional density modes suppress further the on-site fluctuations stabilising in a larger region of phase space the MI state. For $g_{\mathrm{eff}}<g_c$, the density wave instability sets in and competes with dimer order close to $g_c$, $\mathcal{O}_B\approx\mathcal{O}_{DW}$. As the interaction strength is further decreased,  the system condenses into a DW insulator for $t_0/U\ll1$, where $\mathcal{O}_{DW}>\mathcal{O}_{B}$. When $g_{\mathrm{eff}}\ll0$ the system transitions smoothly to the SF phase via an intermediate SS phase.  Bond order is relevant just near the transition to DW ordered states. The intermediate SS phase is suppressed faster as the number of sites increases with respect to the case of Hamiltonian (\ref{JDJBH}).
   
\begin{figure}[t!]
\centering
\includegraphics[width=0.75\textwidth]{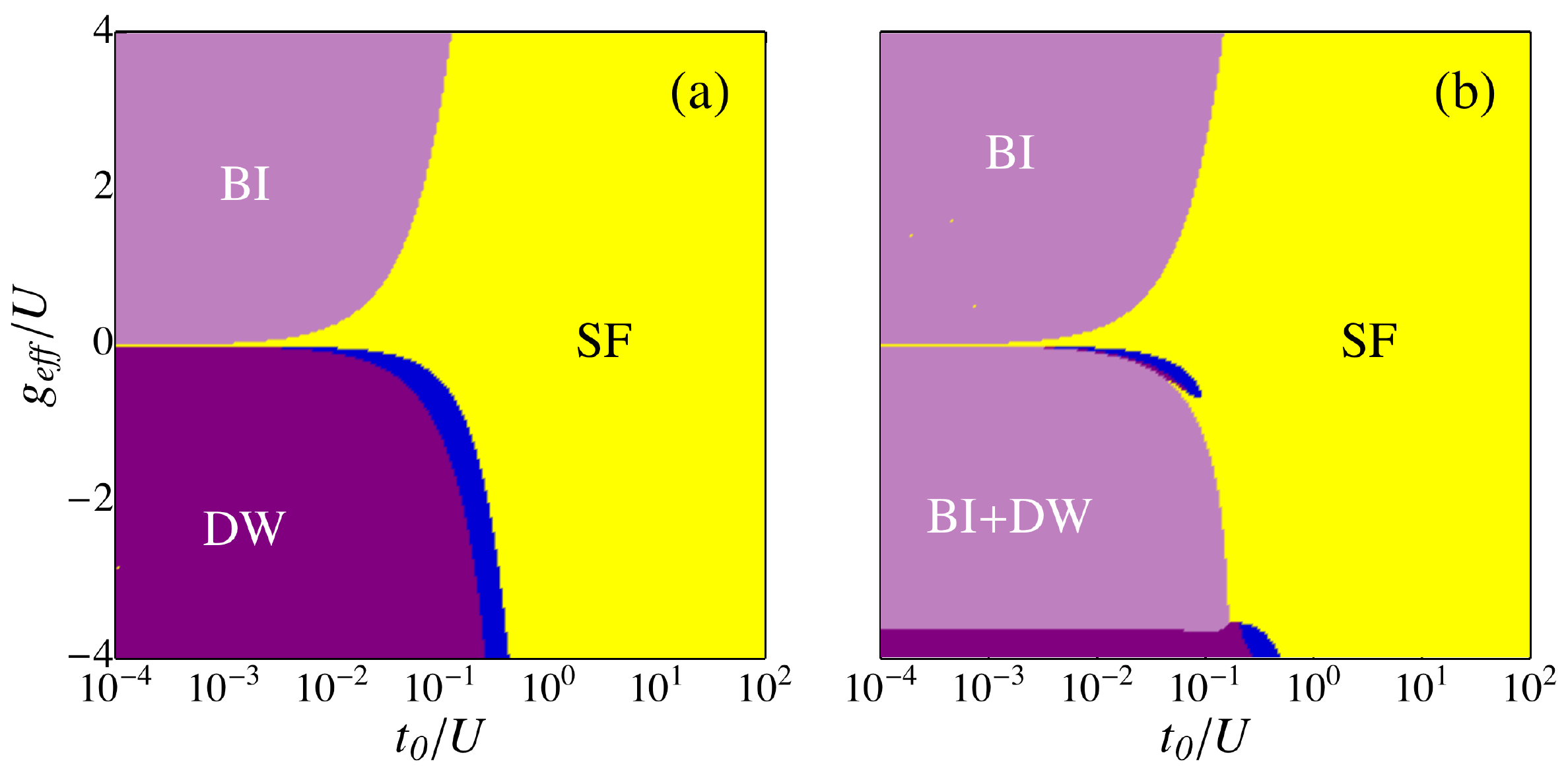}
\captionsetup{width=0.75\textwidth,justification=centerlast,font=small}
\caption{Phase diagrams at half filling $\rho=1/2$ for synthetic interactions via cavity modes. . Panel (a) corresponds to a Bessel like potential. Panel (b)  corresponds to a Morse like potential.  For $g_{\mathrm{eff}}>0$ the phases supported by either potential are qualitatively the same, favouring bond ordered phases for small $t_0/U$. For $g_{\mathrm{eff}}<0$ and $t_0/U\ll1$, DW is the ground state by the Bessel potential (a), while bond order is stabilised in the Morse potential (b) leading to a BI+DW. For both potentials DW appears for $g_{\mathrm{eff}}\ll0$. 
Intermediate SS phases (blue) are shown in both plots.  
Parameters are: $N_s=8$ with $\rho=1/2$.
}
\label{CavitySyn}
\end{figure}

In contrast to unit filling, the half filled case ($\rho=1/2$), offers intriguing behaviour. In the case of light-induced synthetic interactions via cavity modes, we have that the Bessel potential and the Morse potential have different insulating phases for $g_{\mathrm{eff}}\ll0$ and $t_0/U\ll1$, Fig.\ref{CavitySyn}(a) and (b). The Bessel potential supports a DW insulator that smoothly transitions via a SS to a SF state. This is similar to the behaviour seen in the diffraction minima coupling Fig. \ref{PDrho05} (a). However, here the system reaches the SF state and is not SS for large $t_0/U$. In contrast, the Morse potential  supports a BI+DW phase, even though that here there are only density-density interactions. Different from the above, the situation for $g_{\mathrm{eff}}\gg 0$ is qualitatively similar in both potentials, Fig.\ref{CavitySyn}(a) and (b). Here the system for $t_0/U=0$ is in a BI and a continuous transition occurs via  a very narrow intermediate SFD  (not shown) eventually reaching the SF for  $t_0/U\gg0$.

\begin{figure}[t!]
\centering
\includegraphics[width=0.75\textwidth]{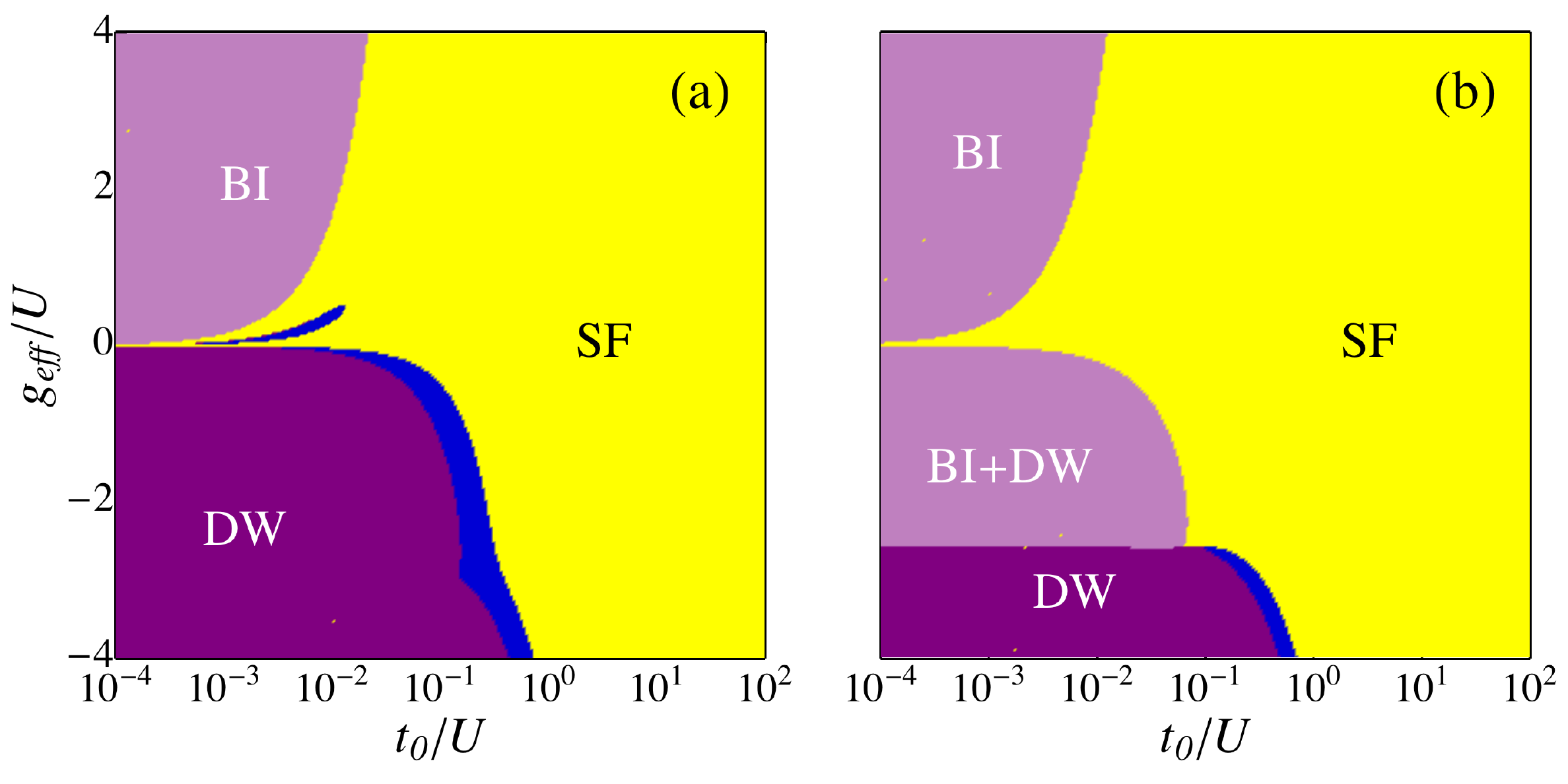}
\captionsetup{width=0.75\textwidth,justification=centerlast,font=small}
\caption{Order parameters at half filling $\rho=1/2$ for synthetic interactions via pump modes as a function of the tunneling $t_0/U$. Panels (a) and (c) correspond to a Bessel like potential. Panels (b) and (d) correspond to Morse like potential. $f_{SF}$ is the superfluid fraction (yellow), $\mathcal{O}_B$ the  bond order parameter (green), $\mathcal{O}_{DW}$ the density wave order parameter (blue) and on-site fluctuations $\Delta(\hat n_i)$ (red). For $g_{\mathrm{eff}}<0$ the phases supported by either potential are qualitatively the same, suppressing bond ordered phases for all $t_0/U$. For $g_{\mathrm{eff}}>0$, bond ordered phases are stabilized for small $t_0/U$. Parameters are: $N_s=8$ with $\rho=1/2$, (a-b): $g_{\mathrm{eff}}/U=-4$ and (c-d): $g_{\mathrm{eff}}/U=+4$.}
\label{PumpSyn}
\end{figure}

When light-induced interactions are constructed via pump modes Fig.\ref{PumpSyn}, a qualitative similarity between phase diagrams for both potentials occurs with respect to Fig.\ref{CavitySyn}. In general, we find suppression of  bond order with respect to the cavity mode case, the transition to the SF occurs at lower critical $t_0/U$. In contrast to the cavity mode case, the pump case is easier to implement experimentally requiring many pump beams compared to a multi-mode cavity or many cavities and the behaviour is qualitatively similar. 

The difference in behaviour between the potentials can be traced to the staggered like nature of the Bessel like potential and the smoother pattern given by the Morse like potential. This combines with the position dependent interaction nature of the many pump/cavity configuration, leading to the effects seen.  In the case of Bessel like potential, even though simulating a finite range, the effect in the system is dominated by the staggered field like nature of its form, alternating coupling between density modes. Thus, the results are similar to density coupling from the single cavity case in the diffraction minima, a staggered density global interaction. In contrast to this, the effect of the Morse like potentials is manifest by stabilising bond ordered phases. Importantly, in the Morse cases Fig.\ref{CavitySyn}(b) and \ref{PumpSyn}(b) one can have BI+DW insulators with density-density interactions. The BI phases for $g_{\mathrm{eff}}>0$ in all the different cases in this section are driven by the quantum fluctuations induced by quantum optical lattice potential, see the previous section. 

\section{Conclusions}

In this paper we have shown the interplay between bond ordered states (bond insulators, supersolid dimers and superfluid dimers),  density wave ordered states (supersolid and density wave insulators), superfluid and Mott insulators due to strong cavity induced interactions in 1D, via exact diagonalization. We have shown numerical evidence to support the identification of bond ordered states with valence bond states, via the calculation of string and parity order parameters in states with bond addressing. We have investigated the suppression and stabilisation of density wave ordered phases and their competition with bond ordered phases due to different choices of synthetic light-induced many-body matter interactions. We have found that using multiple cavity modes and multiple pump modes, one can modify the behaviour of the supported quantum many-body phases in the system. We have found that one can induce bond ordering even by density addressing.  In general, the interplay of the BH model with the cavity induced interaction can change the nature of the quantum phase transitions that appear in the system. These can  be either discontinuous or continuous depending on the design of the spatial profile of the interactions, modifying the typical scenario of the SF-MI transition (BKT type in 1D). We have shown the typical quantum many-body phases the system can support and how do these compete as relevant experimental parameters are changed and different geometries of light are chosen: tunneling, on-site interaction and the effective light induced interaction strength. The interplay between different order parameters demonstrates the connection between the designed light induced interactions and the supported quantum many body-phases in the system. This provides a rich landscape of phases to explore experimentally with intriguing properties. Our results support the possibility to use synthetic many-body matter interactions for the quantum simulation of analogous strongly correlated states related to quantum magnetism from condensed matter. Our findings also suggest, the possibility to use them in the study of the fermionic variant of our system to perform quantum simulations of other interesting states such as, RVB states, which are relevant in mechanisms related to the on set of Hi-temperature superconductivity in real materials. 

Beyond ultracold atoms in cavities, our results can extended to other arrays of naturally occurring, synthetic, hybrid systems or solid state devices with quantum degrees of freedom with strong light-matter coupling in low dimensions. The results in principle can be applied to systems of fermions, spins, molecules (including biological ones)~\cite{LPhys2013}, atoms in multiple cavities~\cite{ArrayPolaritons2006}, 
 ions~\cite{Ions2012} and semiconductor~\cite{SQubits2007}  or superconducting qubits~\cite{JCQubits2009}.  The setup we study might aid in the design of novel quantum materials, where the concepts that we describe can be exploited for the design of properties of real materials and composite devices in solid state systems with strong light-matter coupling. This opens a new chapter on what can be achieved by quantum simulation via atomic systems.

\section*{Acknowledgments}
We would like to thank  D. Hawthorn and F. Piazza for useful discussions. This work was supported by the EPSRC (EP/I004394/1).

  \end{document}